\documentclass[%
 reprint,
 amsmath,amssymb,
 aps,
]{revtex4-2}
\usepackage[utf8]{inputenc}
\usepackage{graphicx}
\usepackage{dcolumn}
\usepackage{orcidlink}
\usepackage{bm}
\usepackage{xcolor}
\usepackage{hyperref}
\begin{document}

\preprint{APS/123-QED}

\title{Effects of charge on the interior volume of BTZ black holes}

\author{Suraj Maurya\orcidlink{0000-0001-6907-8584}}
 \email{p20200471@hyderabad.bits-pilani.ac.in (S. Maurya)}
\author{Sashideep Gutti\orcidlink{0000-0001-7555-8453}}%
 \email{sashideep@hyderabad.bits-pilani.ac.in (S. Gutti)}
 
 \author{Rahul Nigam\orcidlink{0000-0002-0497-5898}}%
 \email{rahul.nigam@hyderabad.bits-pilani.ac.in (R. Nigam)}
\affiliation{Birla Institute of Technology and Science Pilani (Hyderabad Campus)\\ Hyderabad 500078, India}

\date{29-Nov-2022}

\begin{abstract}
In this article we extend the variational technique for maximal volume estimation of a black hole developed by Christodoulou and Rovelli (CR) \cite{CR} to the case of a charged BTZ black hole in 2+1 dimensions. The technique involves a study of the equation of motion of a hypothetical particle moving in an auxiliary manifold defined by spacetime variables. We then compare this estimation with the volume computed using maximization method and extrinsic curvature method. The charge Q of the black hole appears as a log term in the metric and hence an analytical solution for the volume does not exist. So first we compute the steady state radius and the volume for limiting case when the charge Q is very small i.e. $Q << 1$ and then carry out a numerical analysis to solve for the volume for more generic values of the charge. We find that the volume grows monotonically with the advance time. We further investigate the functional behaviour of the entropy of a massless scalar field living on the maximal hypersurface of a near extremal black hole. We show that this volume entropy exhibits a very different functional form compared to the horizon entropy. 
\end{abstract}
\maketitle


\section{Introduction}
A black hole boundary is defined by its event horizon. Event horizons have been studied  exhaustively and they play a crucial role in defining the geometry and more importantly the thermodynamics of a black hole. For a given black hole the position and the area of its event horizon depend on its mass, charge and the angular momentum. So given these  parameters, the area of event horizon is well defined and unique. However computing the volume of a black hole is slightly tricky. In fact there is no unique way of defining the volume of a black hole. A foliation dependent volume is not an unfamiliar concept and we come across this idea even in Minkowski space. However, in whichever way one may define the volume, the boundary area of a black hole which is given by the area of the event horizon remains the same.

There are many interesting ways of defining the black hole volume. Parikh \cite{MP} describes the black hole volume using an invariant slice of spacetime inside the horizon. Gibbons et al. \cite{MG} have offered a thermodynamical definition of the volume i.e. $V_{th}$, which is useful in presence of a cosmological constant $\Lambda$. They define $V_{th}$ as the variable conjugate to the cosmological constant appearing in the first law of thermodynamics for black hole, $dE = TdS + \Omega dJ + \Phi dQ + V_{th} d\Lambda$. The volume definition specific to our interest in this work was provided by Christodoulou and Rovelli (CR) in \cite{CR}. This particular volume grows monotonically with the advance time and has a maximum contribution from a constant $r$ segment. CR demonstrated that computing this volume  is equivalent to finding the geodesic equation of a particle in an equivalent spacetime since the analysis involves extremizing a path length. They showed that this volume for a Schwarzschild black hole increases as $V \sim 3\sqrt{3} m^2 v$, in the limit $v>>m$, where $v$ is the advance time. We used this technique in our earlier work \cite{SSR} and had applied it to BTZ black holes. We showed that the constant $r$ segment of the volume hypersurface can also be derived by demanding a divergence free extrinsic curvature \cite{BR}. In this work we further extend the technique to charged BTZ black holes. Once we obtain the expression for the volume, we probe its thermodynamical aspects using semi-classical approach in the near extremal limit.  

\section{General Setup} \label{EHaction1}
The Hilbert action $(I_H)$ coupled to electromagnetism is defined for charged BTZ black hole as
\begin{equation}\label{action1}
\begin{split}
    I_H = \int d^3x\sqrt{-g}\bigg(\frac{R-2\Lambda}{2k} -\frac{1}{4}F_{\mu\nu}F^{\mu\nu}\bigg) \\
    =\frac{1}{2\pi}\int d^3x\sqrt{-g}\bigg(R + \frac{2}{l^2} -\frac{\pi}{2}F_{\mu\nu}F^{\mu\nu}\bigg)
    \end{split}
\end{equation}
where, $l$ is the AdS length and $\Lambda$ is the cosmological constant which satisfies, $\Lambda = -1/l^2$ and $k=8\pi G/c^4$ is the Einstein gravitational constant. We use natural units where $G = 1/8$ and $\hbar = c = 1$. Hence the factor $1/(16\pi G) = 1/(2\pi)$. The metric of a charged BTZ black hole is defined as  
\begin{equation}\label{metric2}
ds^2 = -N^2(r) dt^2 + \frac{dr^2}{N^2(r)} + r^2 (N^\phi dt + d\phi)^2
\end{equation}
where, shift function $N^\phi = -J/2 r^2$ and the lapse function $N^2(r) = -\Lambda r^2 - M + \frac{J^2}{4 r^2} - \frac{\pi} {2} Q^2 ln( \frac{r}{l})$. Here $M$, $J$ and $Q$ are the mass, angular momentum and the charge carried by black hole, respectively.
The metric above exhibits a coordinate singularity at the zeros of the lapse function so we switch from the Schwarzschild coordinates $(t, r, \phi)$ to Eddington-Finkelstein coordinates $(v, r, \theta)$ to avoid this singularity, which are defined  as
\begin{equation}\label{EFC3}
v = t + \int^r{\frac{dr'}{N^2(r')}} \ and \
\theta = \phi - \int^r{\frac{N^\phi(r')}{N^2(r')}dr'}
\end{equation}
where, $v$ is the advance time. As the cosmological constant, $\Lambda = -1/l^2$ is negative, the background space is AdS. The metric (\ref{metric2}) can now be written as
\begin{multline}\label{CRmetric4}
    ds^2 = -\bigg(-\Lambda r^2 - M - \frac{\pi} {2} Q^2 ln(\frac{r}{l}) \bigg) dv^2 + 2dvdr \\- Jdvd\theta
+ r^2 d\theta^2
\end{multline}
The black hole has an outer and an inner horizon $r_{\pm}^{Q}$. One can define the volume enclosed between the two horizons by picking a spacelike hypersurface. As there is a whole family of such hypersurfaces, the volume enclosed is not unique. We use the CR \cite{CR} technique to define the maximal hypersurface which will enclose maximum volume. 
\section{Analytical solution of the maximum volume}
\subsection{The Maximization technique}
As discussed earlier, in this work we find the hypersurface which encloses the maximum volume inside a black hole. CR \cite{CR} have shown that the maximum contribution to such a hypersurface comes from a constant $r$ segment. We computed such a volume for a BTZ in our earlier work \cite{SSR}. We now use the techniques to understand the effects of charge by studying the interior volume of a charged BTZ black hole.  We will be interested only in the constant $r$ segment whose contribution to the volume of the hypersurface is sufficiently large that we can ignore the remaining segments as justified in \cite{CR}. For the constant $r$ hypersurface, the induced metric (\ref{CRmetric4}) can be written as
\begin{equation}\label{metric5}
ds^2 = -\bigg(-\Lambda r^2 - M - \frac{\pi} {2} Q^2 ln(\frac{r}{l}) \bigg) dv^2  - Jdvd\theta + r^2 d\theta^2
\end{equation}
where, $\Lambda = -1/l^2$. Metric tensor $g_{\mu\nu}$ can be written in the matrix form as
\begin{equation}\label{metrictensor6}
g_{\mu\nu} = 
\Bigg( \begin{matrix}
-f(r) & -J/2 \\  
-J/2 & r^2 
\end{matrix} \Bigg)
\end{equation}
The determinant of $g_{\mu\nu}$ is, $g = det (g_{\mu\nu}) = - r^2 f(r) - \frac{J^2}{4}$, where $f(r) = \bigg(\frac{r^2}{l^2} - M - \frac{\pi} {2} Q^2 ln(\frac{r}{l}) \bigg)$. So the volume for constant $r$ hypersurface is
\begin{equation}\label{volume7}
V_\Sigma  = 2\pi v \sqrt{-r^2 \bigg(\frac{r^2}{l^2} -M - \frac{\pi} {2} Q^2 ln(\frac{r}{l})\bigg)  - \frac{J^2}{4}}
\end{equation}
This is the volume of charged BTZ black hole interior. We seek the radius which we call the steady state radius, $r_{ss}^Q$, which maximizes this volume. Such a steady state radius solves the equation $dV_\Sigma/dr = 0$. So
\begin{equation}\label{volumeprime8}
    \bigg(\frac{dV_\Sigma}{dr}\bigg)_{r= r^Q_{ss}} =  (r^Q_{ss})^2 - \frac{\pi} {4} Q^2 l^2 ln(\frac{r^Q_{ss}}{l})- \frac{M l^2}{2} - \frac{ \pi Q^2 l^2}{8} = 0
\end{equation}
 To get an analytical solution we work in the limit when the charge $Q << 1$. In this limit we expect  $r^Q_{ss} = r_{ss} + \delta$ or $\frac{r^Q_{ss}}{r_{ss}} = 1 + \frac{\delta}{r_{ss}}$,  where $\delta$ is an infinitesimal parameter $( \frac{\delta}{r_{ss}} \ll 1$ or $\frac{r^Q_{ss}}{r_{ss}}-1 \ll 1)$. Substituting the value of $r^Q_{ss}$ in the eq.(\ref{volumeprime8}) and using the approximation on logarithmic function i.e. $\ln(\frac{r^Q_{ss}}{l})\approx \ln(\frac{r_{ss}}{l}) + \frac{\delta}{r_{ss}}$, we get
\begin{equation}\label{eqn9}
    (r_{ss} + \delta)^2 - \frac{\pi} {8} Q^2 l^2 \bigg( 1 + 2 ln(\frac{r_{ss}}{l}) +\frac{2\delta}{r_{ss}}\bigg)- r^2_{ss} = 0
\end{equation}
where, $r_{ss} = l\sqrt{\frac{M}{2}}$, is the steady state radius of BTZ black hole. Solution of equation (\ref{eqn9}) gives,
\begin{equation}\label{delta10}
    \delta = r^Q_{ss} - r_{ss} = \frac{r_{ss} Q^2}{2}\Bigg( \frac{1 + 2ln(\frac{r_{ss}}{l})}{\frac{8 r^2_{ss}}{\pi l^2} - Q^2}\Bigg)
\end{equation}
So we obtain the steady state radius for charged BTZ black hole $(r^Q_{ss})$ from the eq.(\ref{delta10}) as
\begin{multline}\label{steadystateradius11}
    r^Q_{ss} = r_{ss} \Bigg[1 + \frac{Q^2}{2}\Bigg(\frac{1 + 2ln(\frac{r_{ss}}{l})}{\frac{8 r^2_{ss}}{\pi l^2} - Q^2 }\Bigg)\Bigg] \\= l\sqrt{\frac{M}{2}}\Bigg[1 + \frac{Q^2 }{2}\Bigg(\frac{1 + ln(\frac{M}{2})}{\frac{4 M}{\pi} - Q^2 }\Bigg)\Bigg]
\end{multline}
\subsection{Traceless extrinsic curvature technique}
One can arrive at the expression for the steady state radius $r_{ss}$ using another method proposed originally by Bruce L. Reinhart \cite{BR}. We noted that the maximum contribution to the volume is due to the surface r = const. It is argued in \cite{BZ1,BR} that the trace of the extrinsic curvature of such a surface should vanish. So we evaluate the trace of extrinsic curvature of the surface r = cosnt. The unit normal to the surface r = const. for the metric (\ref{CRmetric4}) is given by $n^\alpha = (1/\sqrt{f}, \sqrt{f}, 0)$, where $f(r) = \frac{r^2}{l^2} - M - \frac{\pi}{2}Q^2 \ln(\frac{r}{l}) $. The condition for the vanishing trace of extrinsic curvature is same as making the normal vector divergenceless which is written as $n^\alpha;_{\alpha} = 0$. We note that the volume element of spacetime with metric given by eq.(\ref{CRmetric4}) is $\sqrt{|g|} = r$. So the condition for the vanishing trace of the extrinsic curvature reduces to 
\begin{equation}\label{zerocurvature12}
   n^\alpha;_{\alpha} = \frac{1}{r} \frac{\partial}{\partial r} \bigg(r\sqrt{\frac{r^2}{l^2} - M - \frac{\pi}{2}Q^2 \ln(\frac{r}{l})}\bigg) = 0
\end{equation}
 the eq.(\ref{zerocurvature12}) gives
 \begin{equation}\label{eqn13}
    (r^Q_{ss})^2 - \frac{\pi} {4} Q^2 l^2 ln(\frac{r^Q_{ss}}{l})- \frac{M l^2}{2} - \frac{ \pi Q^2 l^2}{8} = 0
\end{equation}
 which is same as eq.(\ref{volumeprime8}). Hence, the solution of the eq.(\ref{eqn13}) gives same steady state radius $(r^Q_{ss})$ as given in eq.(\ref{steadystateradius11}) i.e.  $r^Q_{ss} = l\sqrt{\frac{M}{2}}\bigg[1 + \frac{Q^2 }{2}\bigg(\frac{1 + ln(\frac{M}{2})}{\frac{4 M}{\pi} - Q^2 }\bigg)\bigg]$. Now, substituting the value of $r^Q_{ss}$ in the eq.(\ref{volume7}) we get the expression for maximum volume as
 \begin{widetext}
\begin{equation}\label{maxvolume14}
    V_\Sigma = \int \sqrt{g} dv d \theta  = 2\pi v \sqrt{-(r^Q_{ss})^2 \bigg(\frac{(r^Q_{ss})^2}{l^2} -M - \frac{\pi} {2} Q^2 ln(\frac{r^Q_{ss}}{l})\bigg)  - \frac{J^2}{4}}
\end{equation}
\end{widetext}
or
\begin{widetext}
\begin{equation}\label{maxvolume15}
  V_\Sigma = 2\pi v \sqrt{- r^2_{ss}\bigg(1 + \frac{\frac{1}{2} + ln(\frac{r_{ss}}{l})}{\frac{8 r^2_{ss}}{\pi Q^2 l^2} - 1}\bigg)^2\Bigg[\frac{r^2_{ss}}{l^2}\bigg(1 + \frac{\frac{1}{2} + ln(\frac{r_{ss}}{l})}{\frac{8 r^2_{ss}}{\pi Q^2 l^2} - 1}\bigg)^2 - M - \frac{\pi}{2}Q^2 ln\frac{r_{ss}}{l}\bigg(1 +  \frac{\frac{1}{2} + ln(\frac{r_{ss}}{l})}{\frac{8 r^2_{ss}}{\pi Q^2 l^2} - 1}\bigg)\Bigg] - \frac{J^2}{4}}
\end{equation}
\end{widetext}
Note here $r_{ss} =l \sqrt{\frac{M}{2}}$, which is the steady state radius in absence of the charge $Q$ and the expression for volume simplifies to $V_\Sigma = \pi v \sqrt{M^2l^2 -J^2}$ when the charge is taken to zero. The region plot of $V_{\Sigma}$ from eq.(\ref{maxvolume15}) as shown below. Here we take the AdS length $l = 4$.
\begin{figure}[htp]
    \centering
    \includegraphics[width=.9\linewidth]{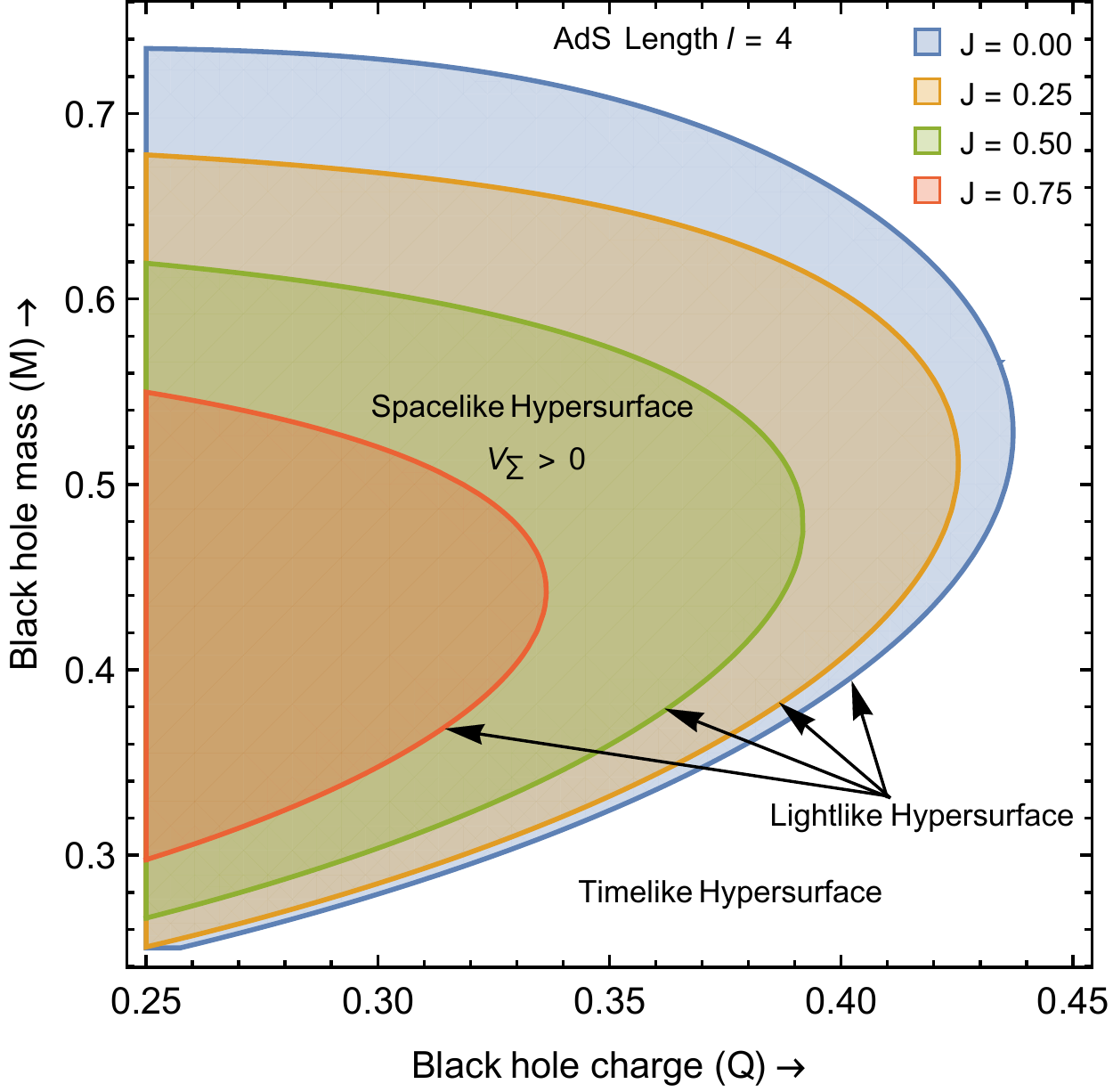}
    \caption{\small{ The graph highlights the region in the parameter space spanned by the mass $M$, charge $Q$ and angular momentum $J$ of a charged BTZ black hole. The different shaded regions correspond to angular momentum $(J = 0, 0.25, 0.50, 0.75)$. The volume of hypersurface $V_\Sigma$ of black hole is positive definite and hence well defined inside the shaded regions. In the region outside, the hypersurface becomes timelike and cannot be used to define the black hole volume. The boundary of the regions corresponds to lightlike hypersurface which again cannot be used to define the volume. Here we take the AdS length $l=4$.}}
    \label{fig:1}
\end{figure}\\
  The volume of this hypersurface defining the black hole volume increases monotonically with the advanced time $v$. As discussed by many \cite{BZ1,BZ2,MZ,BR1,SSR}, such a behaviour can be appropriately used in defining and understanding the black hole entropy. We will touch on this in latter sections. Note that the expression for $V_{\Sigma}$ also provides a constraint on the values of the charge $Q$ and angular momentum $J$ for the $V_{\Sigma}$ to exist. A negative value for the expression inside the square root would mean the volume becomes timelike and hence cannot be used to define the black hole volume. Demanding that the expression inside the square root is positive definite, we find the region in the parameter space spanned by $Q$ and $J$, which we call the permitted region for $V_{\Sigma}$ to exist. Note that these graphs in the figure (\ref{fig:1}) become less accurate for larger values of $Q$ as we are working with approximation that $Q << 1$.
\subsection{Variational method in the auxiliary manifold}
In this section we use the formulation developed by Christodoulou et al. \cite{CR} to construct the spacelike hypersurface $\Sigma$ and check if it reconciles with our results in previous section. We write $\Sigma$, which is a 2-manifold, as a direct product of a 1-sphere defined by $\theta$ and a curve $\gamma$ embedded in the 2-dimensional space $(v,r)$ paramterized by $\lambda$ as
\begin{equation}
\begin{split}
\Sigma \sim \gamma \times S^1 \\
\gamma \sim (v(\lambda), r(\lambda))
\end{split}
\end{equation} 
The line element of the induced metric on the hypersurface $\Sigma$ is
\begin{equation}\label{inducemetric17}
ds^2_{\Sigma} = \big(-f(r)\dot{v}^2 + 2\dot{v}\dot{r} \big)d\lambda^2 - J\dot{v} d\lambda d\theta  + r^2 d\theta^2
\end{equation}
The dot represents derivative with respect to the parameter $\lambda$ and $f(r) = \frac{r^2}{l^2} -M - \frac{\pi}{2}Q^2 \ln(\frac{r}{l})$. For $\Sigma$ to be a spacelike surface, we require
\begin{equation}\label{eqn18}
r^2\bigg(-f(r)\dot{v}^2 + 2\dot{v}\dot{r}\bigg) - \frac{J^2\dot{v}^2}{4} > 0
\end{equation}
and the proper volume of $\Sigma$ is given by 
\begin{equation} \label{propervolume19}
V_{\Sigma}[\gamma] = 2 \pi\int_0^{\lambda_f} d\lambda \sqrt{r^2\bigg(-f(r)\dot{v}^2 + 2\dot{v}\dot{r}\bigg) - \frac{J^2\dot{v}^2}{4}} 
\end{equation}
This volume depends on the curve $\gamma$ and we want to find the curve $\gamma$ which would maximize this volume. This can be viewed as an extremization problem and our goal then is to find the equations of motion for the Lagrangian 
\begin{equation}\label{lagrangian20}
L(r,\dot{r},v,\dot{v}) = \sqrt{r^2\bigg(-f(r)\dot{v}^2 + 2\dot{v}\dot{r}\bigg) - \frac{J^2\dot{v}^2}{4}} 
\end{equation}
We will find the equations of motion and then use appropriate normalization to set 
\begin{equation}\label{eqn21}
L(r,\dot{r},v,\dot{v}) = 1
\end{equation}
This normalization ensures that $\Sigma$ is spacelike. 
The curve $\gamma$ parameterized by $\lambda$ has two end points which we choose to be at $r(\lambda_i = 0) = r_+^Q$ and $r(\lambda_f) =r^Q_-$, so the curve stretches between the inner and the outer horizon and $v(\lambda_i = 0) = v$ and $v(\lambda_f) = v_f$.

\subsubsection{\textbf{Euler-Lagrange equations of motion}}
The Lagrangian is a function of the variables $v$ and $r$. First we write the Euler-Lagrangian equation corresponding to $v$ as
\begin{equation}\label{eqn22}
\frac{\partial L}{\partial \dot{v}} = \frac{1}{2L}\Bigg(r^2 (-2f(r)\dot{v}+ 2\dot{r}) -\frac{J^2\dot{v}}{2}\Bigg) = a
\end{equation}
using eqs.(\ref{lagrangian20}) and (\ref{eqn21}) this simplifies to 
\begin{eqnarray}\label{vdot23}
&&\dot{v} \bigg( -2r^2f(r) - J^2/2 \bigg) = 2a - 2\dot{r} r^2 \nonumber \\ 
&&\Rightarrow \dot{v} = \frac{\dot{r}r^2 - a}{r^2 f(r) + J^2/4} 
\end{eqnarray}
Next using eqs.(\ref{eqn22}) and (\ref{vdot23}), we get
\begin{equation}\label{vdot24}
\dot{v} = \frac{1}{a + \dot{r}r^2} 
\end{equation}
and plugging this expression in eq.(\ref{lagrangian20}), we find
\begin{equation}\label{rdot25}
\dot{r} = - \frac{1}{r^2}\sqrt{r^2f(r) + \frac{J^2}{4} + a^2}    
\end{equation}
The expression for $\dot{v}$ is same as deduced by others for the case of BTZ black hole and the expression for $\dot{r}$ reduces to the expression for BTZ black hole as shown by \cite{MZ} when we put $J$ and $Q = 0$. 
\subsubsection{\textbf{Steady state radius and maximum volume}}
As we discussed earlier, most of the contribution to the volume of the maximal hypersurface comes from a steady state segment which corresponds to a constant $r_{ss}$. We get the expression for $r^Q_{ss}$ by taking $\dot{r} = 0$ and equating the right hand side of eq.(\ref{rdot25}) to zero. 
\begin{multline}\label{eqn26}
(r^Q_{ss})^2 \bigg(\frac{(r^Q_{ss})^2}{l^2} -M - \frac{\pi} {2} Q^2 ln(\frac{r^Q_{ss}}{l})\bigg)  + \frac{J^2}{4} + a^2 = 0\\
\Rightarrow a = \sqrt{-(r^Q_{ss})^2 \bigg(\frac{(r^Q_{ss})^2}{l^2} -M - \frac{\pi} {2} Q^2 ln(\frac{r^Q_{ss}}{l})\bigg)  - \frac{J^2}{4}}
\end{multline}
Note that for $\dot{r} = 0$, we have  $\dot{v} = 1/a$. As $v$ must be a monotonically increasing function of the parameter $\lambda$, $a$ must be positive definite. Now the volume of the hypersurface as defined in eq.(\ref{propervolume19}) can be evaluated as
\begin{equation}\label{volume27}
    V_{\Sigma} = 2\pi \int d\lambda = 2\pi a \int dv = 2\pi a v
\end{equation}
Substituting the expression for $a$ as in eq.(\ref{volume27}), we get 
\begin{equation}\label{volume28}
V_{\Sigma} = 2\pi v \sqrt{-(r^Q_{ss})^2 \bigg(\frac{(r^Q_{ss})^2}{l^2} -M - \frac{\pi} {2} Q^2 ln(\frac{r^Q_{ss}}{l})\bigg)  - \frac{J^2}{4}}
\end{equation}
Further substituting the expression for $r^Q_{ss}$ obtained from the traceless extrinsic curvature technique, we obtain
\begin{widetext}
\begin{equation}\label{volume29}
  V_\Sigma = 2\pi v \sqrt{- r^2_{ss}\bigg(1 + \frac{\frac{1}{2} + ln(\frac{r_{ss}}{l})}{\frac{8 r^2_{ss}}{\pi Q^2 l^2} - 1}\bigg)^2\Bigg[\frac{r^2_{ss}}{l^2}\bigg(1 + \frac{\frac{1}{2} + ln(\frac{r_{ss}}{l})}{\frac{8 r^2_{ss}}{\pi Q^2 l^2} - 1}\bigg)^2 - M - \frac{\pi}{2}Q^2 ln\frac{r_{ss}}{l}\bigg(1 +  \frac{\frac{1}{2} + ln(\frac{r_{ss}}{l})}{\frac{8 r^2_{ss}}{\pi Q^2 l^2} - 1}\bigg)\Bigg] - \frac{J^2}{4}}
\end{equation}
\end{widetext}
This matches with the expression for the volume we derived earlier using the maximization technique. Hence, the variational method provides a different way to calculate the volume of a black hole using Euler-Lagrange equation of motion of a hypothetical particle which is traversing along a geodesic curve in the auxiliary manifold.
\section{Numerical solution of the maximum volume}
\begin{figure}[htp]
    \centering
    \includegraphics[width=.9\linewidth]{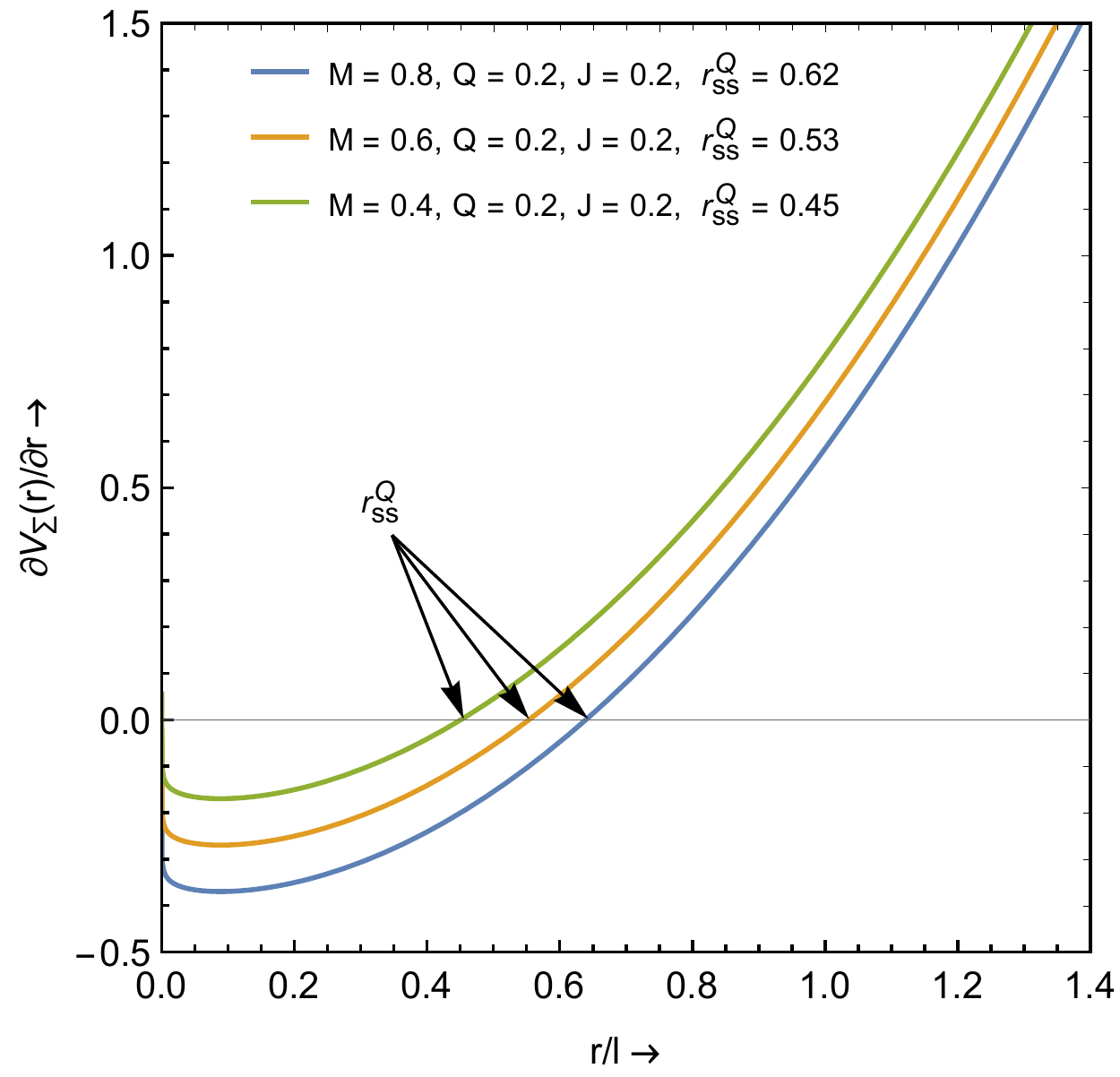}
    \caption{\small{The roots of this graph give the steady state radii $(r^Q_{ss})$ of a charged BTZ black hole at different values of black hole parameters $(M, Q, J)$. Here we take the AdS length $l=1$.}}
    \label{fig:2}
\end{figure}
 Our analysis till now has been limited to the case when the black hole charge is very small i.e. $Q << 1$. This was necessary as solving the equation analytically for the volume with the $ln(r^Q_{ss}/l)$ term in eq.(\ref{maxvolume14}) is not possible. However the $Q <<1$ limit provides a good insight into effects of adding charge on the volume $V_{\Sigma}$ and the steady state radius. Moreover as we will study a certain thermodynamical aspects of a charged BTZ black hole in later sections, we will note that such an investigation is possible only in small charge limit regime. In this section we carry out a numerical analysis to study the behaviour of the steady state radius $r_{ss}$ and subsequently of the maximal hypersurface by removing any restrictions on the values of charge $Q$. The steady state radius $r_{ss}$ is defined by the maximization equation $dV_\Sigma/dr = 0$. So we plot $dV_\Sigma/dr$ versus $r/l$ seeking the values of $r$ for which for function has its zeros. The plot is shown in figure (2), here we take the AdS length $l=1$. \\\\
Once we have the values of the the steady state radii $r_{ss}$ for different mass $M$, charge $Q$ and angular momentum $J$ from the graph, we plug the values into the expression from eq.(\ref{maxvolume15}) and evaluate $V_\Sigma$. Here we tabulate the volumes of charged BTZ and BTZ black hole for a comparison. In the case of charged BTZ black hole, the volume depends on the charge Q and evidently its numerical value is smaller than the volume of an BTZ black hole. The complete table is shown below:
\\
\\
\textbf{Table:} Maximum volume $(V_{\Sigma{max}})$ of charged-BTZ black hole and BTZ black hole at different value of steady state radius $(r^Q_{ss})$ is given in the following table. Here we are taking AdS length $l = 1$. 
\begin{widetext}
\scalebox{1.25}{\begin{tabular}{c c c c c c c}
    \hline
    Mass  & Charge  & Angular Momentum &charged-BTZ & charged-BTZ &  charged-BTZ & BTZ  \\
    $M$ & $Q$ & $J$ & $r^Q_{ss}$ &$r^Q_+$&  ${(V_{\Sigma{max}})}/{2 \pi v}$ &  ${(V_{\Sigma{max}})}/{2 \pi v}$\\
    \hline
    0.4&0.2&0.2&0.45&0.47&0.11&0.18\\
    
     0.5&0.2&0.2&0.47&0.67&0.20&0.23\\
    
    0.6&0.2&0.2&0.53&0.75&0.26&0.29\\
    
      0.7&0.2&0.2&0.57&0.82&0.32&0.34\\
     
     0.8&0.2&0.2&0.62&0.88&0.37&0.39\\
    
    0.9&0.2&0.2&0.66&0.94&0.42&0.44\\
    \hline
\end{tabular}}
\end{widetext}
From the above table we note that the event horizon is always larger than the steady state radius, $r^Q_+ > r^Q_{ss}$. This means that the steady state radius $r^Q_{ss}$ is always hidden behind the event horizon $r^Q_{+}$. By adding a small charge $Q$ we find that the volume of charged BTZ black hole decreases and less than the BTZ black hole.

\section{Analytical solution of the horizons}
\begin{figure}[htp]
    \centering
    \includegraphics[width=0.9\linewidth]{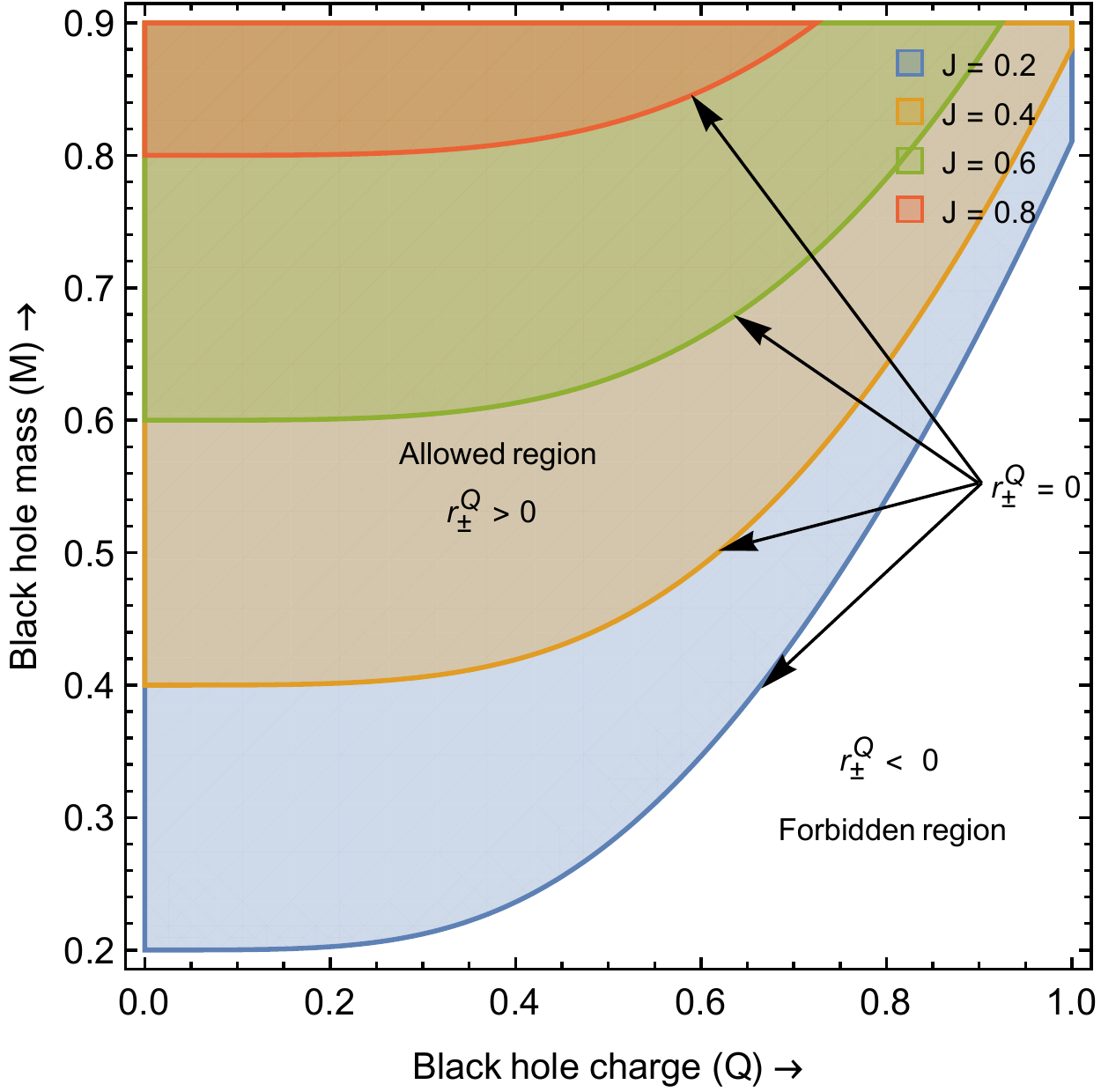}
    \caption{\small{Graph between Mass $(M)$ and charge $(Q)$ of charged BTZ black hole for different values of the angular momentum J. The shaded regions represent the parameters values which allow the existence of horizons and unshaded regions correspond to the parameter values which forbid horizons to exist. Here we take the AdS length $l=1$}}
    \label{fig:3}
\end{figure}

The horizons of a charged BTZ black hole are the roots of the lapse function $N^2(r)$ which is defined as
\begin{equation}\label{lapse30}
\begin{split}
    N^2(r) = \frac{r^2}{l^2} - M + \frac{J^2}{4 r^2} - \frac{\pi} {2} Q^2 ln(\frac{r}{l})
\end{split}
\end{equation}
The outer and the inner horizons are given by the zeros of $[N^2(r)]_{r = r^Q_{\pm}} = 0$.
Again, due to the presence of the  log term, the equation does not have an analytical solution and we rely on perturbative approach in the limit when charge $Q \rightarrow 0$. In this limit the horizons are only slightly displaced from the horizons of the BTZ black hole \cite{SSR} given by $r_{\pm}^2 =\frac{Ml^2}{2}\big(1\pm \sqrt{1-(\frac{J}{Ml})^2}\big)$. Let $r^Q_{\pm} = r_{\pm} + \delta$ or $\frac{r^Q_{\pm}}{r_{\pm}} = 1 + \frac{\delta}{r_{\pm}}$, where $\delta$ is an infinitesimal parameter such that $\frac{\delta}{r_{\pm}} \ll 1$ or $\frac{r^Q_{\pm}}{r_{\pm}} - 1 \ll 1$. So $\log(\frac{r^Q_{\pm}}{l})\approx\log(\frac{r_{\pm}}{l}) + \frac{\delta}{r_{\pm}}$. With this approximation, the horizon equation can be written as 
\begin{equation}\label{delta31}
  \frac{(r_{\pm} + \delta)^2}{l^2} - M + \frac{J^2}{4 (r_{\pm} + \delta)^2} - \frac{\pi} {2} Q^2 \bigg(ln(\frac{r_{\pm}}{l}) +\frac{\delta}{r_{\pm}}\bigg) = 0
\end{equation}
Solving eq.(\ref{delta31}) to the first order in $\delta$ gives
\begin{equation}\label{delta32}
    \delta = r^Q_{\pm} - r_{\pm} = \frac{\pi Q^2 r_{\pm} ln(\frac{r_{\pm}}{l})}{4\bigg(\frac{r^2_{\pm}}{l^2} - \frac{J^2}{r^2_{\pm}} - \frac{\pi Q^2}{4}\bigg)} 
\end{equation}
Hence, the horizons of a charged BTZ black hole $(r^Q_{\pm})$ are 
\begin{multline}\label{CReventhor33}
    r^Q_{\pm} = r_{\pm} \Bigg[ 1+ \frac{\pi Q^2 ln(\frac{r_{\pm}}{l})}{4\bigg(\frac{r^2_{\pm}}{l^2} - \frac{J^2}{r^2_{\pm}} - \frac{\pi Q^2}{4}\bigg)} \Bigg]\\
    = r_{\pm}\Bigg[1+\frac{\pi Q^2 ln(\frac{r_{\pm}}{l})}{4\bigg(M\sqrt{1- (\frac{J}{Ml})^2}-\frac{\pi Q^2}{4}\bigg)} \Bigg]
\end{multline}
 As we take the limit $Q <<1$, the denominator in the above expression is positive definite and the term $ln(r_{\pm}/l)$ in the numerator is negative as $r_{\pm} < l$. So, the sign of $\delta$ in the eq.(\ref{delta32}) is negative and hence $r_{\pm}^Q < r_\pm $. We note that the position of the horizons of the charged BTZ with respect to the neutral one does not depend on the sign but the magnitude of the charge $Q$. The region plot of event horizon shown in figure (\ref{fig:3}).
\section{Numerical solution of the horizons}
\begin{figure}[hbp]
    \centering
    \includegraphics[width=0.9\linewidth]{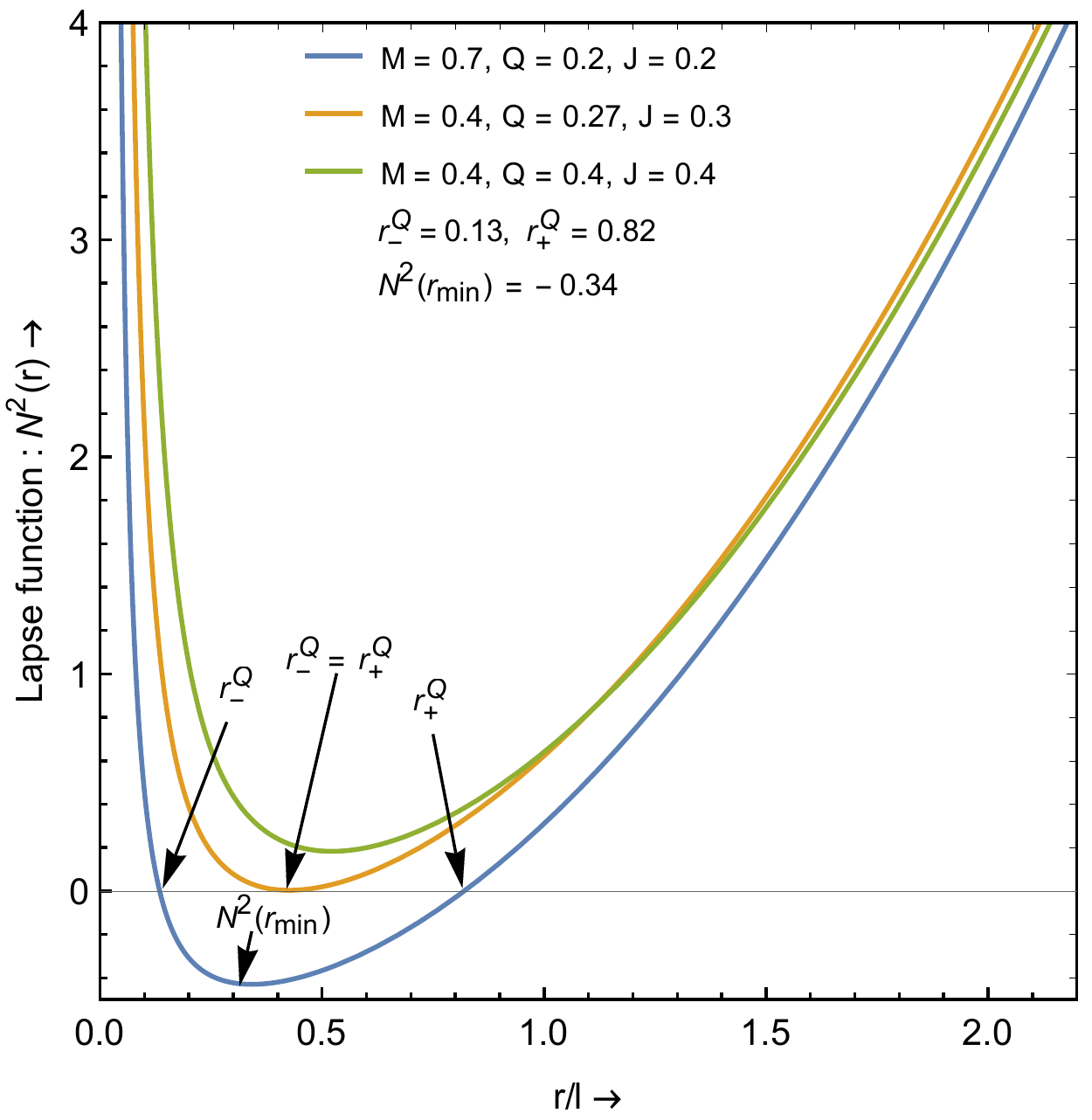}
    \caption{\small{Lapse function $N^2(r)$ versus $r/l$ for different values of black hole parameters $(M,Q,J)$. Blue curve depicts the existence of two different horizons, the red curve shows the case when both the horizons overlap and green curve depicts the case when there is no horizon.}}
    \label{fig:4}
\end{figure}

\begin{figure}[htp]
    \centering
    \includegraphics[width=0.9\linewidth]{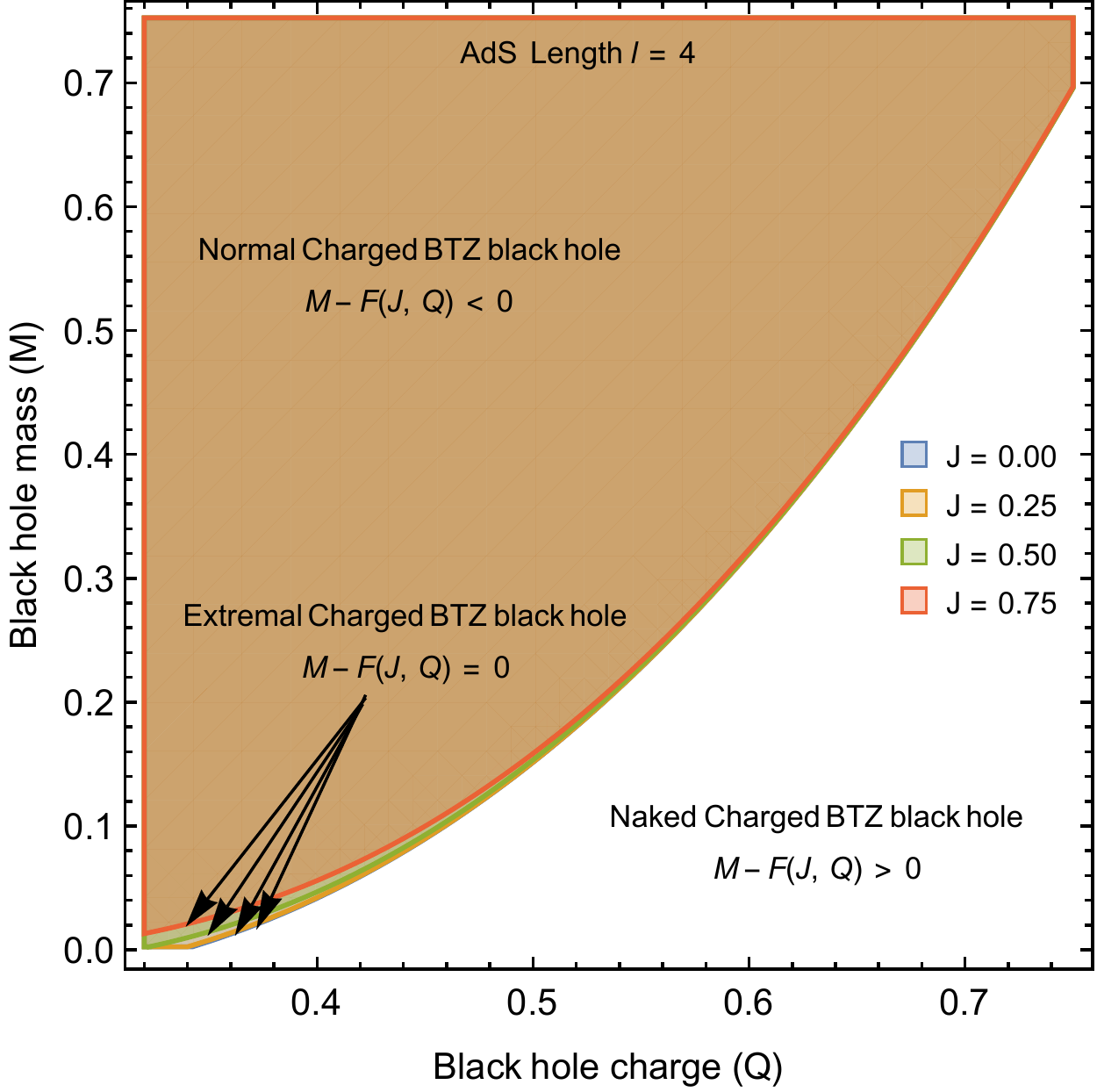}
    \caption{\small{ Graph between black hole mass $M$ and charge $Q$ for different values of the angular momentum $J$. Here the shaded regions represent a normal charged BTZ black hole with well defined horizons, the unshaded region represent a naked charged BTZ black hole and the boundary between these two regions represents the extremal charged BTZ case.}}
    \label{fig:5}
\end{figure}

Now we investigate the parameter space spanned by mass $M$, charge $Q$ and angular momentum $J$ to understand how the black hole horizons vary over different regions. The existence of the horizons is guaranteed only when the lapse function has real zeros. Generally the lapse function has two roots which correspond to the two horizons. But, for a certain values of the charges, the lapse function may have only one root and that would correspond to an extremal black hole. In this section we analyse these regions by imposing different conditions on the black hole parameters. We illustrate these cases in the figure (\ref{fig:4}) by taking different values of the black hole parameters to get all the possible outcomes. Let us consider the case of an extremal black hole which is obtained when the lapse function has only one root. This is possible only when the minima of the lapse function touches the $r/l$-axis in $N^2(r)- r/l$ plot. We solve the equation, $\big(\frac{d [N^2(r)]}{dr}\big)_{r = r_0} = 0$, to get the value of  $r^2_0$ as 
\begin{equation}\label{eqn34}
r^2_0 = \frac{l^2}{8} \bigg(\pi Q^2 + \sqrt{\pi^2 Q^4 + \frac{16 J^2}{l^2} } \bigg)
\end{equation}. 
We check that at $r = r_0$, $\frac{d^2 [N^2(r)]}{dr^2} > 0$, so $r = r_0 = r_{min}$ and the polynomial $N^2(r)$ has a minima at $r_0$. Hence, the minimum value of $N^2(r)$ is
\begin{equation}\label{lapsemin34}
N^2(r_{min}) = \frac{r^2_{min}}{l^2} - M + \frac{J^2}{4 r^2_{min}} - \frac{\pi} {2} Q^2 ln(\frac{r_{min}}{l})
\end{equation}
Plugging the expression for $r_{min}$ from eq.(\ref{eqn34}), we get 
\begin{multline}\label{lapsemin35}
    N^2(r_{min}) = - M + \frac{\pi Q^2 + \sqrt{\pi^2 Q^4 + \frac{16 J^2}{l^2} }}{8}\\ + \frac{2 J^2}{l^2 \bigg(\sqrt{\pi^2 Q^4 + \frac{16 J^2}{l^2}}\bigg)}\\ - \frac{\pi}{4} Q^2 ln\bigg[ \frac{1}{8} \bigg(\pi Q^2 + \sqrt{\pi^2 Q^4 + \frac{16 J^2}{l^2} } \bigg) \bigg] = -M + F(J,Q)
\end{multline}
where, $F(J,Q) = \frac{\pi Q^2 + \sqrt{\pi^2 Q^4 + \frac{16 J^2}{l^2} }}{8} + \frac{2 J^2}{l^2 \bigg(\sqrt{\pi^2 Q^4 + \frac{16 J^2}{l^2}}\bigg)} - \frac{\pi}{4} Q^2 ln\bigg[ \frac{1}{8} \bigg(\pi Q^2 + \sqrt{\pi^2 Q^4 + \frac{16 J^2}{l^2} } \bigg) \bigg]$\\
The graph between $N^2(r)$ vs. $r/l$ as shown below. Note that we take the AdS length $l = 1$ here. 
\\
There exist three possible situations which is arising from the eq.(\ref{lapsemin35}) and we investigate them one at a time. \\
\textbf{Case (a).} When $N^2(r_{min}) < 0$ : In this case both the roots of the polynomial $N^2(r)$ are real and distinct i.e. $r = r^Q_\pm$. This is the normal charged BTZ black hole. This is the general condition for the existence of two horizons of a charged BTZ black hole. If charge Q = 0, then we get $M > J/l$ or $J < Ml$, which is the normal case of a BTZ black hole. So, from the eq.(\ref{lapsemin35}) for $N^2(r_{min})<0$, we obtain an inequality relation between mass $M$ and polynomial $F(J,Q)$ as
\begin{equation}\label{eqn36}
 M - F(J,Q) >0
\end{equation}
\textbf{Case (b).} When $N^2(r_{min}) = 0$ : In this case both root of the polynomial $N^2(r)$ are real and equal i.e.  $r = r_- = r_+$. This is the extremal case of a charged BTZ black hole in which both horizons coincide. If charge Q = 0 then, we get $M = J/l$ or $J = Ml$, which is the extremal case for a BTZ black hole. So, from the eq.(\ref{lapsemin35}) for $N^2(r_{min})= 0$, we get the relation between mass $M$ and the polynomial $F(J,Q)$ as
\begin{equation}\label{eqn37}
 M -  F(J,Q) = 0
\end{equation} 
\textbf{Case (c).} When $N^2(r_{min}) > 0$ : In this case both the roots of the polynomial $N^2(r)$ are imaginary and hence the horizons do not exist. This is the case of naked singularity. If charge Q = 0, then we get the inequality $M < J/l$ or $J > Ml$, which is the case of naked singularity of a BTZ black hole. So, from the eq.(\ref{lapsemin35}) for $N^2(r_{min})>0$ we get the inequality relation between mass $M$ and polynomial $F(J,Q)$ as
\begin{equation}\label{eqn38}
 M - F(J,Q) < 0
\end{equation}
The graphical representation of the nature of black hole for all three possible cases as discussed above is depicted in the figure (\ref{fig:5}).
\section{Entropy of a massless scalar field in the black hole volume}
After understanding the geometry of spacelike hypersurfaces inside the charged BTZ black hole, we now try to gain some insight into the process of state counting in the interior of a black hole. Since a black hole evaporates and shrinks with time through Hawking radiation, it is not in an equilibrium state. Moreover the metric inside a black hole is not static. In general we cannot do a state counting in such a dynamical background. However we can avoid this problem by considering a near extremal black hole which has the horizon temperature close to absolute zero. One can indeed apply the techniques of quantum statistics to do a state counting in such a quasi-static background and we work in this limit through the rest of the analysis. Now we count the accessible states in the phase space \cite{BC}. The 4-dimensional phase space for a charged BTZ black hole is spanned by the conjugate variables $v,\theta,p_v$ and $p_{\theta}$. One quantum state corresponds to a cell of volume $(2\pi\hbar)^2$ in this 4-dimensional phase space. The total number of quantum states would be given by the integral of the measure 
\begin{equation}
    \frac{dvd\theta dp_v dp_{\theta}}{(2\pi\hbar)^2}
\end{equation}
over the accessible region. To find the accessible region, we introduce a massless scalar field on the hypersurface $\Sigma$ which satisfies the Klein-Gordon equation,  $\frac{1}{\sqrt{g}}\partial_{\mu}(\sqrt{g}g^{\mu\nu} \partial_{\nu} \phi) = 0$. As we focus on the part of $\Sigma$ which corresponds to $\dot{r} = 0$, the line element as given in eq.(\ref{inducemetric17}) would reduce to $ds^2 = -f(r^Q_{ss})dv^2 - Jdvd\theta + (r^Q_{ss})^2 d\theta^2$. Hence, the metric tensor $g_{\mu\nu}$ define as 
\begin{equation}\label{metric39}
g_{\mu\nu} = \Bigg( \begin{matrix}
-f(r^Q_{ss})&-\frac{J}{2}\\  
\frac{-J}{2}&(r^Q_{ss})^2 
\end{matrix} \Bigg)
\end{equation}
Let $\lambda_v$ and $\lambda_{\theta}$ be the eigenvalues of the above metric $g_{\mu\nu}$. So, we get $\lambda_v$$\lambda_{\theta}$ = g = det$(g_{\mu\nu})$. The Klein-Gordon equation in the momentum space gives the following relation 
\begin{equation}\label{KG40}
E^2 = \frac{1}{\lambda_v} p_v^2 + \frac{1}{\lambda_{\theta}} p_\theta^2
\end{equation}
Now we count the number of states in the phase space with energy less than E. That would be written as 
\begin{multline}\label{energystates41}
g(E) = \frac{1}{(2\pi\hbar)^2}\int \int dvd\theta dp_v dp_\theta \\= \frac{1}{(2\pi\hbar)^2}\int\ dvd\theta \int_{-E/\sqrt{\lambda_{\theta}}}^{E/\sqrt{\lambda_{\theta}}} dp_{\theta}\bigg[\int_{- \sqrt{(E^2 - \lambda_{\theta}p_\theta^2)/\lambda_{\theta}}}^{\sqrt{(E^2 - \lambda_{\theta}p_\theta^2)/\lambda_v}}dp_v \bigg]\\  =\frac{\pi E^2}{(2\pi\hbar)^2}\int \sqrt{\lambda_v\lambda_{\theta}} dv d\theta =\frac{\pi E^2}{(2\pi\hbar)^2}\int \sqrt{g} dv d\theta
 =\frac{ E^2V_{\Sigma}}{4\pi}
\end{multline}
where, we put $\hbar = 1$. The limits of both the integrals are $P_v =\pm \sqrt{(E^2 - \lambda_{\theta}p_\theta^2)/\lambda_v}$ and $p_{\theta}=\pm \sqrt{E/\lambda_{\theta}}$. Now the free energy for the scalar field at inverse temperature $\beta = (k_BT)^{-1}$ is defined as
\begin{multline}\label{freeenergy42}
     F(\beta) = \frac{1}{\beta} \int dg(E) ln (1 - e^{-\beta E})\\
     = \frac{1}{\beta}\int\frac{g(E)(-\beta)e^{-\beta E}dE}{1-e^{-\beta E}}
     = -\int \frac{g(E)dE}{e^{\beta E}-1}
\end{multline}
From eqs.(\ref{energystates41}) and (\ref{freeenergy42}), we get
\begin{equation}\label{freeenergy44}
    F(\beta)  = -\frac{v}{2} \int \frac{E^2 dE}{e^{\beta E} -1} = -\frac{\zeta(3)  V_{\Sigma}}{2\pi\beta^3}
\end{equation}
where, $\zeta(3)$ is the Riemann zeta function. We then obtain the entropy of a scalar field living on the maximal hypersurface $\Sigma$ corresponding to maximal slicing as
\begin{equation} \label{entropy46}
     S_\Sigma = -\frac{\partial F}{\partial T} = k_B\beta^2 \frac{\partial F}{\partial \beta} = \frac{3\zeta(3)V_{\Sigma}}{2\pi\beta^2}\propto \frac{v}{\beta^2}
\end{equation}
where, we takes the Boltzmann constant $(k_B)=1$ in final expression. The eq.(\ref{entropy46}) shows that the entropy of the maximal hypersurface $\Sigma$ corresponding to maximal slicing grows linearly with advance time $(v)$. The inverse temperature of the field here can be well approximated by the horizon temperature $T_H$ in the near extremal limit. Now the horizon temperature of a black hole is defined as 
\begin{equation}\label{horiztempr46}
    T_H = \bigg(\frac{\partial S_H}{\partial M}\bigg)_{J,Q}^{-1}
\end{equation}
where, $S_H$ is the horizon entropy of the black hole. We now find the relation between the advance time, the field temperature and the black hole parameters as follows. 

In 1972, Jacob Bekenstein \cite{JB} proposed that black hole should have entropy which is proportional to the area of the horizon. In 1974 Stephen Hawking \cite{SWH,SH} used quantum field theoretical calculations in curved space time to show that black holes emit thermal radiation corresponding to a certain temperature. Further, using the thermodynamic relationship between energy, temperature and entropy, Hawking was able to confirm Bekenstein's proposal and fix the constant of proportionality at 1/4. The horizon entropy is also known as Bekenstein-Hawking entropy which is proportional to surface area of the black hole. For the case of 2+1 dimensional BTZ black hole,  the horizon entropy $S_H$ is found to be two times the horizon length $L_H$ \cite{BTZ} i.e.
\begin{equation}\label{horizonentropy47}
    S_H = \frac{2L_H}{l_p} =  \frac{4\pi r^Q_+}{l_p} 
\end{equation}
where, $l_p = \sqrt{\hbar G/c^3}$ is the Planck length and $r^Q_+$ is the event horizon of the charged BTZ black hole. Substituting the value of $r^Q_+$ from the eq.(\ref{CReventhor33}) we get the horizon entropy as
\begin{multline}\label{horizentropy48}
      S_H = \frac{4\pi r_{+}}{l_p} \Bigg[1+\frac{\pi Q^2 ln(\frac{r_+}{l})}{4(\frac{r^2_+}{l^2} - \frac{J^2}{4r^2_+} - \frac{\pi Q^2}{4})}\Bigg]\\
    = \frac{4\pi r_{+}}{l_p}\Bigg[1+\frac{\pi Q^2 ln(\frac{r_+}{l})}{4\bigg(M\sqrt{1- (\frac{J}{Ml})^2}-\frac{\pi Q^2}{4}\bigg)} \Bigg]
\end{multline}
We proceed by plugging eq.(\ref{horizonentropy47}) into the expression for horizon temperature of charged BTZ black hole as defined in eq.(\ref{horiztempr46}) to get
\begin{equation}\label{horiztempr49}
    T_H = \bigg(\frac{\partial S_H}{\partial M}\bigg)_{J,Q}^{-1}= 
    \frac{1}{4\pi} \bigg(\frac{\partial r^Q_+}{\partial M}\bigg)_{J,Q}^{-1} 
\end{equation}
Now differentiating the horizon radius $r_+^Q$ with respect to the black hole mass $M$ we get
\begin{widetext}
    \begin{multline}\label{eqn50}
       \bigg(\frac{\partial r^Q_+}{\partial M}\bigg)_{J,Q}  =\frac{\partial r_+}{\partial M}\Bigg[\frac{\bigg(\frac{r^2_+}{l^2} -\frac{J^2}{4r^2_+} - \frac{\pi Q^2}{4} \bigg)^2 + \frac{\pi Q^2}{4}\Bigg \{\bigg(\frac{r^2_+}{l^2} -\frac{J^2}{4r^2_+} - \frac{\pi Q^2}{4} \bigg) -\bigg(\frac{r^2_+}{l^2} +\frac{3J^2}{4r^2_+}- \frac{\pi Q^2}{4} \bigg)ln\bigg(\frac{r_+}{l}\bigg)\Bigg\} }{\bigg(\frac{r^2_+}{l^2} -\frac{J^2}{4r^2_+} - \frac{\pi Q^2}{4} \bigg)^2}\Bigg]\\
   =\frac{\partial r_+}{\partial M}\Bigg[\frac{\bigg(M\sqrt{1-(\frac{J}{Ml})^2}-\frac{\pi Q^2}{4}\bigg)^2  + \frac{\pi Q^2}{4}\Bigg \{\bigg(M\sqrt{1-(\frac{J}{Ml})^2}-\frac{\pi Q^2}{4}\bigg) -\bigg(2M -M\sqrt{1 - (\frac{J}{Ml})^2} - \frac{\pi Q^2}{4}\bigg) ln(\frac{r_+}{l})\Bigg\} }{\bigg(M\sqrt{1-(\frac{J}{Ml})^2}-\frac{\pi Q^2}{4}\bigg)^2}\Bigg]
\end{multline}
\end{widetext}
We use the expression for the event horizon $r_+$ of a BTZ black hole in 2+1 dimensions \cite{SSR,BTZ} to get  
\begin{widetext}
\begin{equation}\label{rotatinghoriz51}
    r^2_+ = \frac{Ml^2}{2}\Bigg(1+ \sqrt{1-(\frac{J}{Ml})^2}\Bigg)
    \Rightarrow \frac{\partial r_+}{\partial M} = \frac{r_+}{2M\sqrt{1-(\frac{J}{Ml})^2}}=\frac{r_+ l^2}{2(r^2_+ - r^2_-)}
\end{equation}
\end{widetext}
then using the eqs.(\ref{horiztempr49}), (\ref{eqn50}) and (\ref{rotatinghoriz51}), we get the horizon temperature $(T_H)$ for charged BTZ black hole as 
\begin{widetext}
\begin{equation}\label{horiztempr52}
    \begin{split}
      T_H = \frac{(r^2_+ - r^2_-)}{2\pi r_+l^2}\Bigg[\frac{\bigg(\frac{r^2_+}{l^2} -\frac{J^2}{4r^2_+} - \frac{\pi Q^2}{4} \bigg)^2}{\bigg(\frac{r^2_+}{l^2} -\frac{J^2}{4r^2_+} - \frac{\pi Q^2}{4} \bigg)^2 + \frac{\pi Q^2}{4}\Bigg \{\bigg(\frac{r^2_+}{l^2} -\frac{J^2}{4r^2_+} - \frac{\pi Q^2}{4} \bigg) -\bigg(\frac{r^2_+}{l^2} +\frac{3J^2}{4r^2_+}- \frac{\pi Q^2}{4} \bigg)ln\bigg(\frac{r_+}{l}\bigg)\Bigg\} }\Bigg]\\
      =\frac{\frac{M}{2\pi r_+}\sqrt{1-(\frac{J}{Ml})^2}\bigg(M\sqrt{1-(\frac{J}{Ml})^2}-\frac{\pi Q^2}{4}\bigg)^2}{\bigg(M\sqrt{1-(\frac{J}{Ml})^2}-\frac{\pi Q^2}{4}\bigg)^2 + \frac{\pi Q^2}{4}\bigg[\bigg(M\sqrt{1-(\frac{J}{Ml})^2}-\frac{\pi Q^2}{4}\bigg) - \bigg(2M -M\sqrt{1 - (\frac{J}{Ml})^2} - \frac{\pi Q^2}{4}\bigg) ln(\frac{r_+}{l}) \bigg]}
      \end{split}
\end{equation}
\end{widetext} 
As discussed earlier, we need to approach the extremal limit for the statistical counting to make sense. In the near extremal limit the Hawking temperature tends to zero and a quasi static limit is attained. For a BTZ black hole, this limit is attained as the mass tends to the angular momentum $(Ml \rightarrow J)$. In our case of a charged BTZ, we further have to make the charge $Q$ very small to get $T_H \rightarrow 0$. So at the near extremal limit, $Ml- J = \delta = lQ^2$, from eq.(\ref{horiztempr52}) we get the horizon temperature to be $T_H \propto -\frac{1}{l} \frac{\sqrt{\delta/l}}{ln(r_+/l)} \propto \frac{1}{l} \frac{\sqrt{\delta/l}}{|ln(r_+/l)|}$.
Note that the horizon $r_+$ is smaller than the AdS length $l$ so, $ln(r_+/l)$ is a negative quantity. The expression for the Hawking temperature is positive definite and tends to zero in the near extremal limit. Now as the Hawking radiation is thermal in nature, the mass loss of the black hole per unit horizon area can be written using the Stefan-Boltzmann law for 2+1 dimensions which is defined as 
\begin{equation}\label{Stefan-Bolzmann53}
    \frac{1}{r^Q_+}\frac{dM}{dv} = -\sigma T_H^3
\end{equation}
where, $\sigma = 5.670374419\times 10^{-8}\frac{W}{m^2 K^4} $ is the  Stefan-Boltzmann constant. We want to write the advance time $v$ as a function of the black hole parameters so we integrate this equation in the near extremal limit in which the event horizon of charged BTZ black hole becomes $r^Q_+ = l\sqrt{\frac{M}{2}}\bigg[1 + \frac{\pi}{4} \sqrt{\frac{\delta}{2Ml}}ln\bigg(\sqrt{\frac{r_+}{l}}\bigg)\bigg] \approx l\sqrt{\frac{M}{2}}$. Now eq.(\ref{Stefan-Bolzmann53}) can be written as
\begin{multline}\label{vprop54}
     v = -\frac{1}{\sigma}\int \frac{dM}{r^Q_+ T^3_H} \approx \frac{\sqrt{2}l^2}{\sigma}\bigg(\sqrt{\frac{l}{\delta}}\bigg)^3 \int \frac{[ ln(\sqrt{M/2})]^3}{\sqrt{M}}dM
\end{multline}
after solving the integral (\ref{vprop54}) we get
\begin{equation}\label{vprop55}
     v \approx \frac{6\sqrt{2}l^2\sqrt{M}[ ln(\frac{r_+}{l})]^2}{\sigma (\sqrt{\delta/l})^3 } \bigg[1 -\frac{1}{3} ln(\frac{r_+}{l})- 2\bigg(\frac{1- ln(\frac{r_+}{l})}{ [ln(\frac{r_+}{l})]^2}\bigg)\bigg]
\end{equation}
Here, $\beta = (k_B T)^{-1} = (k_B T_H)^{-1} \propto l\frac{ |ln(\frac{r_+}{l})|}{k_B\sqrt{\delta/l}}$. Now dividing the eq.(\ref{vprop55}) by $\beta^2$, we get
\begin{equation}\label{eqn56}
   \frac{v}{\beta^2} \approx \frac{6\sqrt{2}\sqrt{M}k^2_B}{\sigma}\sqrt{\frac{l}{\delta}}\bigg[1 -\frac{1}{3} ln(\frac{r_+}{l})- 2\bigg(\frac{1- ln(\frac{r_+}{l})}{ [ln(\frac{r_+}{l})]^2}\bigg)\bigg]
\end{equation}
 Hence, from the eqs.(\ref{entropy46}) and (\ref{eqn56}), we get the entropy of the scalar field living on the maximal hypersurface $\Sigma$ to be
\begin{equation}\label{eqn57}
    S_{\Sigma} \propto  \frac{r_+}{l}\bigg[1 - \frac{1}{3}ln(\frac{r_+}{l})- 2\bigg(\frac{1- ln(\frac{r_+}{l})}{ [ln(\frac{r_+}{l})]^2}\bigg)\bigg]
\end{equation}
 In this section we calculated the volume entropy using a semi-classical approach. As the horizon entropy of the black hole is proportional to the horizon length i.e. $S_H\propto r_+$, evidently the two expressions do not match. The second and the third terms inside the bracket in eq.(\ref{eqn57}) are not small compared to 1 and hence cannot be thought of as corrections. In fact if we try to make the second term small by choosing an appropriate ratio $r_+/l$, the third term becomes large and vice versa. We note that the two terms together are less than $0.5$ only if $r_+/l < 0.1$ and $S_H$ and $S_\Sigma$ are of same order. However, in general the formula for the internal entropy $S_\Sigma$ exhibits quite a different functional behavior compare to the horizon entropy $S_H$. This is a different result from our earlier work \cite{SSR}, where we found that both $S_\Sigma$ and $S_H$ were proportional to each other. Other earlier works \cite{BZ1,BZ2,MZ,BR1,BR2} have also found the two entropies to be proportional to each other. But all these results were restricted to neutral black holes. Evidently the different functional form of entropy is a result of introducing charge $Q$ in the metric. It will be interesting to understand the exact physical reason for this difference which we plan to investigate in our future work.
 
\section{Conclusions}
In this work we have extended the idea of maximal slicing proposed by Bruce L. Reinhart \cite{BR} and indendently by F. Estabrook \cite{FE} and the variational method to setup the Euler-Lagrange equation of motion by Christodoulou and Rovelli \cite{CR} for Schwarzschild black hole, to the case of a charged BTZ black hole. It is also an extension of our earlier work in which we had probed a BTZ black hole \cite{SSR}. The charge appears as a logarithmic term in the metric which makes it difficult to find an analytical solution. So a part of our work is done in the small charge limit. We have computed the steady state radius and the volume of the maximal spacelike hypersurface in this limit. The expression for the volume allowed us to impose constraints on the black hole parameters mass $M$, Charge $Q$ and angular momentum $J$. We further investigated the system numerically and also analysed the locations of the black hole horizons with respect to the steady state radius. We found the conditions on the black hole parameters for which the horizons exist which is necessary for the maximal slicing technique to be valid. We then investigated a certain thermodynamical aspects of the maximal volume. We do this by introducing a massless scalar field on the maximal hypersurface and computing the entropy in the near extremal limit of black hole. The near extremal limit was necessary to justify a static background in which quantum statics could be applied. We found the expression for the volume entropy $S_\Sigma$ as a function of horizon length. This volume entropy $S_\Sigma$ is found to exhibit a very different functional behaviour compared to the Bekenstein-Hawking horizon entropy $S_H$. It will be interesting to study this different nature of the two entropy functions.      

\begin{acknowledgments}
  We would like to thank our institute BITS Pilani Hyderabad campus for providing the required infrastructure to carry out this research work. Suraj Maurya would further like to thank Government of India for providing  CSIR (NET-JRF) fellowship to support this research work.
\end{acknowledgments}

\end{document}